\documentclass{desyproc}

\usepackage{subfigure}

\begin{document}
\title{\vspace{-3cm}
\hfill{\small{IPPP/08/57; DCPT/08/114}}\\[2cm]
 WISP hunting -- some new experimental ideas}

\author{{\slshape Joerg Jaeckel}\\[1ex]
Institute for Particle Physics and Phenomenology, Durham University, Durham DH1 3LE, UK\\
}

\contribID{jaeckel\_joerg}

\desyproc{DESY-PROC-2008-02}
\acronym{Patras 2008} 

\maketitle

\begin{abstract}
We present several new ideas on how to search for weakly interacting
sub-eV particles in laboratory experiments. The first experiment is
sensitive to minicharged particles. It exploits that in
strong electric fields particle - antiparticle pairs are produced by the Schwinger mechanism.
The charged particles move
along the lines of the electric field and generate a current that
can be measured. The other two experiments are designed to search for
hidden-sector photons. They are based on photon - hidden photon
oscillations and resemble classic light shining through a wall
experiments. One uses (nearly) constant magnetic fields instead of the laser light.
Photon - hidden photon mixing would allow these magnetic fields to
leak through superconducting shielding which would ordinarily
eliminate all magnetic fields. The other
one replaces the laser light with microwaves
inside cavities. The latter can achieve much higher quality factors
than optical cavities increasing the sensitivity.
\end{abstract}

\section{Introduction}
Over the last few years it became increasingly clear that low energy
experiments can provide a powerful tool to explore hidden
sectors of particles which interact only very weakly with the ordinary standard model particles.
Such hidden sectors appear in many extensions of the standard model. In fact, it may be exactly those hidden sectors that give us crucial information
on how the standard model is embedded into a more fundamental theory as, e.g., string theory.

The key observation from the viewpoint of low energy experiments is that, due to their feeble interactions with the standard model particles, the hidden sector particles
are relatively unconstrained allowing them to be light possibly even in the sub-eV range. This opens the possibility for observable effects in low energy but high precision
experiments.

In this note we will focus on two particular classes of such light `hidden-sector' particles: minicharged particles and hidden sector photons.
The former are particles interacting with the ordinary electromagnetic field via the usual minimal coupling induced by the covariant derivative,
\begin{equation}
D_{\mu}=\partial_{\mu}-{\mathbf{i}}Q_{f}eA_{\mu}
\end{equation}
where $Q_{f}$ is the electric charge of the particle of a particle $f$.
For example if $f$ is a fermion the interaction term reads
$Q_{f}e \bar{f}A\!\!\! / f.$

The crucial point for a minicharged particle is now simply that the charge is much smaller than~$1$,
\begin{equation}
Q_{f}\ll 1.
\end{equation}
In particular it is not necessarily integer. Indeed it does not even have to be a rational number.
Minicharges can arise in theories with kinetic mixing \cite{Holdom:1985ag} (see also below) but also in scenarios with extra dimensions \cite{Batell:2005wa}.
Typical predicted values, e.g., in realistic string compactifications range from $10^{-16}$ to $10^{-2}$ \cite{Batell:2005wa,Dienes:1996zr}.

The second class of particles we are concerned with in this note are massive hidden sector photons.
These are extra U(1) gauge bosons which can mix with the ordinary electromagnetic photons via a so-called kinetic mixing term~\cite{Holdom:1985ag} in the Lagrangian,
\begin{equation}
\label{lagrangian}
{\mathcal{L}}= -\frac{1}{4} F^{\mu\nu}F_{\mu\nu}-\frac{1}{4}X^{\mu\nu}X_{\mu\nu}
-\frac{1}{2}\chi\,F^{\mu\nu}X_{\mu\nu}  +\frac{1}{2}m^{2}_{\gamma^{\prime}} X_\mu X^\mu+j_\mu A^\mu,
\end{equation}
where $F_{\mu\nu}$ is the field strength tensor for the ordinary
electromagnetic {U(1)$_{_\mathrm{QED}}$} gauge field~$A^{\mu}$,
$j^\mu$ is its associated current (generated by
electrons, etc.) and $X^{\mu\nu}$ is the
field strength for the hidden-sector {U(1)$_\mathrm{h}$} field
$X^{\mu}$. The first two terms are the standard kinetic terms for
the photon and hidden photon fields, respectively. Because the field
strength itself is gauge invariant for U(1) gauge fields, the third
term is also allowed by gauge and Lorentz symmetry.  This term
corresponds to a non-diagonal kinetic term, the kinetic
mixing~\cite{Holdom:1985ag}. This term is a renormalizable dimension
four term and does not suffer from mass suppressions. It is
therefore a sensitive probe for physics at very high energy scales.
Kinetic mixing arises in field theoretic \cite{Holdom:1985ag} as
well as in string theoretic setups
\cite{Batell:2005wa,Dienes:1996zr} and
predictions for its size range between $10^{-16}$ and $10^{-2}$. The second
to last term is a mass term for the hidden photon. This could either
arise from a Higgs mechanism or it could be a St\"uckelberg mass
term~\cite{Stueckelberg:1938}.

\section{AC/DC an experiment to search for minicharged particles}
\begin{figure}[t]
\begin{center}
\includegraphics[angle=0,width=.45\textwidth]{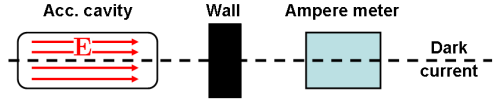}
\end{center}
\vspace{-0.6cm} \caption{\small
Schematic illustration of an
\emph{accelerator cavity dark current} (AC/DC) experiment
 for searching minicharged particles.}
\label{acdc}
\end{figure}

The basic setup \cite{Gies:2006hv} is depicted in Fig. \ref{acdc}.
In a strong electric field a vacuum pair of charged particles gains energy if the particles are separated by a distance along the lines of the electric field.
If the electric field is strong enough (or the distance large enough) the energy gain can overcome the rest mass, i.e. the virtual particles turn into real particles.
This is the famous Schwinger pair production mechanism~\cite{Schwinger:1951nm}.
After their production the electric field accelerates the particles and antiparticles according to their charge in opposite directions.
This leads to an electric current (dashed line in Fig.~\ref{acdc}).
If the current is made up of minicharged particles the individual particles have very small charges and interact
only very weakly with ordinary matter. Therefore, they can pass even through thick walls nearly unhindered. An electron current, however, would be stopped.
After passage through the wall we can then place an ampere meter to detect the minicharged particle current.

Typical accelerator cavities achieve field strengths of $\gtrsim 25\,{\rm MeV/m}$ and their size is typically of the order of 10s of cm.
Precision ampere meters can certainly measure currents as small as $\mu {\rm A}$ and even smaller currents of the order of ${\rm pA}$ seem feasible. Using the Schwinger pair production
rate we can then estimate the expected sensitivity for such an experiment to be
\begin{equation}
\epsilon_{\rm sensitivity} \sim 10^{-8}-10^{-6}\quad{\rm for}\,\,m_{\epsilon}\lesssim {\rm meV}.
\end{equation}
Therefore such an experiment has the potential for significant improvement over the currently best
laboratory\footnote{Astrophysical bounds are much stronger~\cite{Davidson:2000hf} but are also somewhat model
dependent~\cite{Masso:2006id}.} bounds~\cite{Gies:2006ca,Badertscher:2006fm}, $\epsilon\lesssim{\rm few}\,\, 10^{-7}$.

\section{Searching hidden photons inside a superconducting box}

\begin{figure}
\begin{center}
\subfigure{
\includegraphics[width=0.45\linewidth]{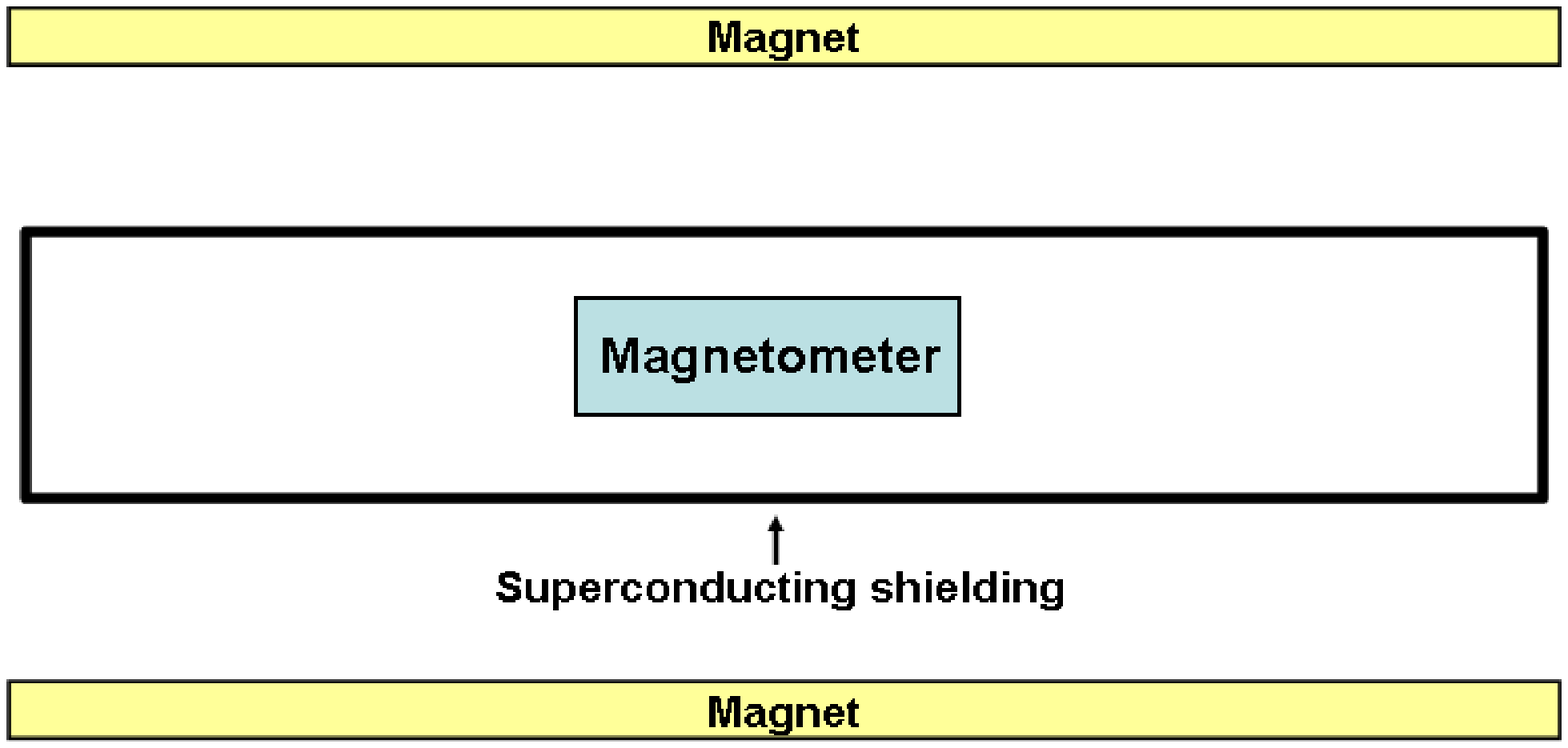}}
\hspace*{0.5cm}
\subfigure{
\includegraphics*[width=0.45\linewidth,bb=13 20 1163 515]{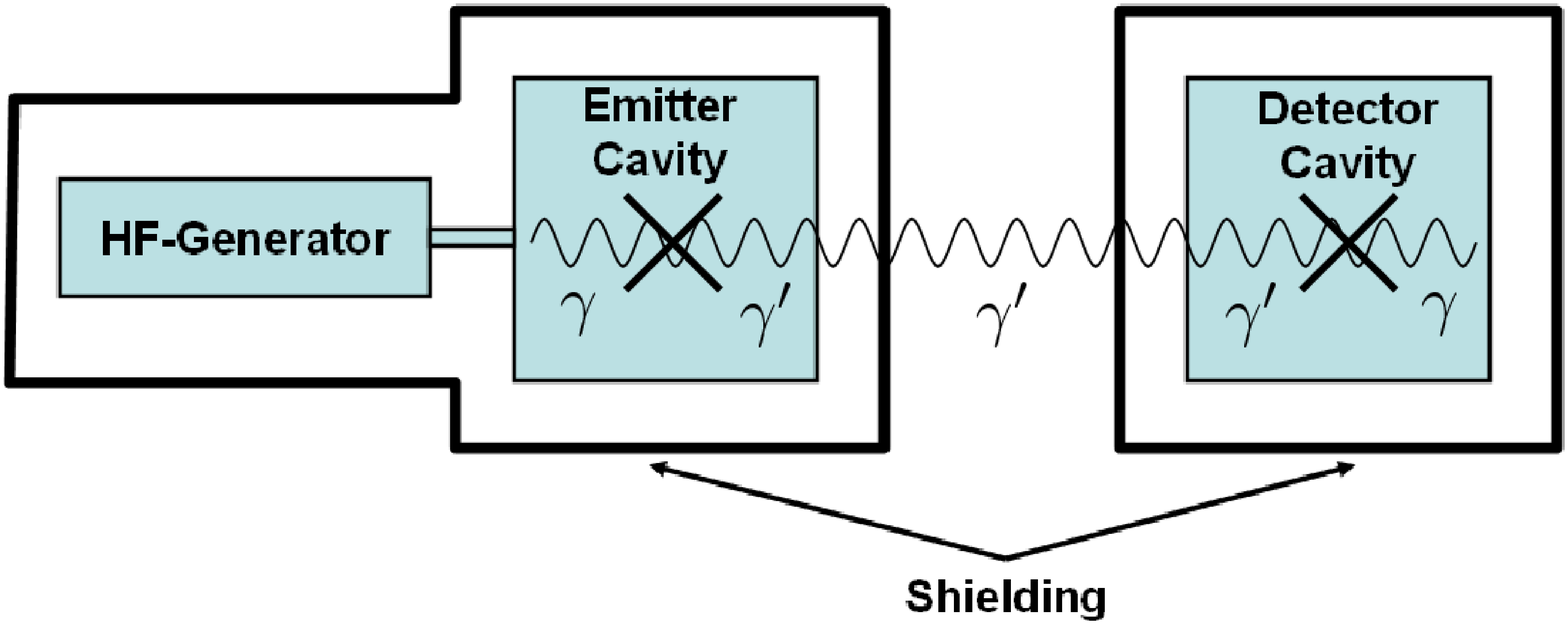}
}
\end{center}
\vspace{-4ex}
\caption[...]{\small
{\bf Left panel:} Sketched setup for the \emph{superconducting box} experiment.
{\bf Right panel:} Schematic illustration of a
\emph{microwaves permeating through a shielding} experiment
 for the search for massive hidden sector photons mixing
with the photon  (a high-frequency
(HF) generator drives the emitter cavity).
\label{magnet}}
\end{figure}

The basic idea \cite{Jaeckel:2008sz} of the proposed experiment is very similar to a classic \emph{light shining through a wall experiment}~\cite{regeneration}.
However, instead of light it uses a static magnetic field and the wall is replaced by superconducting shielding (cf. the left panel of Fig. \ref{magnet}).
Outside the shielding we have a strong magnetic field. Upon entering the superconductor the ordinary electromagnetic field
is exponentially damped with a length scale given by the London penetration depth $\lambda_{\rm Lon}$. Yet, due to the photon -- hidden photon mixing
a small part of the magnetic field is converted into a hidden magnetic field.
After the superconducting shield is crossed the mixing turns a small fraction of the hidden magnetic field back into an ordinary magnetic field that can be
detected by a magnetometer.
Since the magnetometer measures directly the field (and not some probability or power output) the signal is proportional
to the transition amplitude and therefore to the mixing squared, $\chi^2$, instead of being proportional to $\chi^4$.

High precision magnetometers can measure
fields of the order of $10^{-13}$~T and even tiny fields of a few~$10^{-18}$~T seem feasible.
The expected sensitivity is shown as the blue area in Fig.~\ref{limits}.

\section{A cavity experiment to search for hidden photons}

\begin{figure}
\begin{center}
\includegraphics[width=0.45\linewidth]{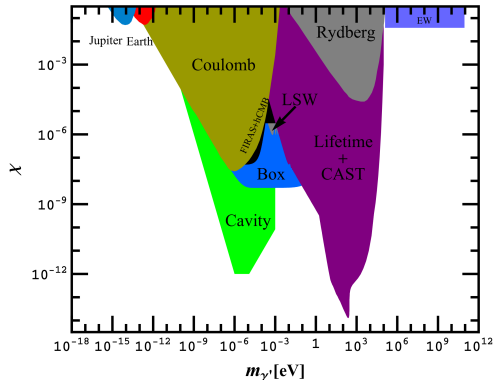}
\end{center}
\vspace{-5ex}
\caption[...]{\small
Current bounds on hidden-sector photons (cf., e.g., \cite{Ahlers:2008qc} and references therein).
The \emph{superconducting box} experiment could probe the blue region (Box). The estimated sensitivity for the \emph{microwaves permeating through a shielding}
is shown in green (Cavity). For details on the respective setups see \cite{Jaeckel:2008sz,Jaeckel:2007ch}.
\label{limits}}
\end{figure}

Our final proposal \cite{Jaeckel:2007ch} (see
\cite{Hoogeveen:1992uk} for a similar proposal for axions
\footnote{The only change necessary for an axion search is that one
applies an additional magnetic field which allows for the usual
photon--axion conversion inside magnetic fields. One might be
worried that
in this case one cannot use superconducting cavities because a magnetic field applied from outside the cavity cannot permeate through the superconductor to the
inside of the cavity where it is needed for the conversion. This would allow only normal conducting cavities which have somewhat smaller $Q$.
However, this may not be the case if one uses type II superconductors which allow
for magnetic field penetration (via flux tubes) while maintaining their superconducting properties. Nevertheless, the magnetic field (and the flux tubes) can
increase the surface resistance, again limiting the $Q$ factor. Further investigation is needed to determine if one can achieve high $Q$ with a strong magnetic field on
the inside of the cavity.})
is another setup
searching for signatures of photon -- hidden photon oscillations
which resembles a classic \emph{light shining through a wall}~\cite{regeneration} experiment, more precisely a
resonant setup~\cite{Sikivie:2007qm}. It consists of two microwave cavities
shielded from each other (cf. right panel of Fig.~\ref{magnet}).
In one cavity, hidden photons are produced
via photon -- hidden photon oscillations. The second, resonant, cavity is then driven by the hidden photons that
permeate the shielding and reconvert into photons. Due to the high quality factors achievable for
microwave cavities (superconducting ones can reach $Q\sim 10^{11}$) and the good sensitivity of microwave detectors $\sim 10^{-26}-10^{-20}$~W such a setup will allow for
an unprecedented discovery potential for hidden sector photons in the mass range  from
$\mu{\rm{eV}}$ to ${\rm{meV}}$ (green area in Fig.~\ref{limits}).

\section{Conclusions}
We have presented several ideas for small scale laboratory experiments to search for weakly interacting sub-eV particles predicted
in many extensions of the standard model. For minicharged particles an
\emph{accelerator cavity/dark current} experiment promises improvement over current laboratory bounds. Both the \emph{superconducting box} and
the \emph{microwaves permeating through a shielding} experiment have the potential to improve not only upon the current laboratory but also beyond
existing astrophysical and cosmological bounds, thereby having significant discovery potential for new physics. Searching for extremely weakly interacting
particles at small masses that would be missed in conventional colliders all these experiments provide for a new, complementary probe of fundamental physics.

Finally, we would like to point out that an experiment of the \emph{microwaves permeating through a shielding} type is
already in an initial stage~\cite{Penny} and will also be used to search for axions and axion-like particles.

\section*{Acknowledgements}
The author wishes to thank the organizers of the \emph{4th Patras Workshop on Axions, WIMPs and WISPs} for a very enjoyable and stimulating meeting.
Moreover he is indebted to M.~Ahlers, H.~Gies, J.~Redondo and A.~Ringwald for many many interesting discussions and fruitful collaboration.

\begin{footnotesize}

\end{footnotesize}


\end{document}